\documentclass[review]{elsarticle}

\usepackage{multirow,setspace,amssymb,amsmath,graphicx,color,rotating,subfigure,url}

\usepackage{natbib}
\usepackage{textcomp}
\usepackage{CJK}
\usepackage{bm}%
\usepackage{rotating}
\usepackage{epsfig}
\usepackage{dcolumn}
\usepackage{epstopdf}
\newtheorem{theorem}{Theorem}

\begin{document}

\begin{frontmatter}

\title{Information Spreading Dynamics on Adaptive Social Networks}

\author[inst1]{Chuang Liu}
\author[inst1]{Nan Zhou}
\author[inst2]{Xiu-Xiu Zhan}\ead{x.zhan@tudelft.nl}  
\author[inst3,inst4]{Gui-Quan Sun}\ead{gquansun@126.com}  
\author[inst1]{Zi-Ke Zhang}\ead{zhangzike@gmail.com}  

\address[inst1]{Alibaba Research Center for Complexity Sciences, Hangzhou Normal University, Hangzhou, 311121, P. R. China}
\address[inst2]{Faculty of Electrical Engineering, Mathematics and Computer Science, Delft University of Technology, 2628 CD Delft, The Netherlands}
\address[inst3]{Department of Mathematics, North University of China, Taiyuan 030051, Shanxi, People's Republic of China}
\address[inst4]{Complex Systems Research Center, Shanxi University, Taiyuan 030006, Shanxi, People's Republic of China}

\begin{abstract}
There is currently growing interest in modeling the information diffusion on social networks across multi-disciplines. The majority of the corresponding research has focused on information diffusion independently, ignoring the network evolution in the diffusion process. Therefore, it is more reasonable to describe the real diffusion systems by the co-evolution between network topologies and information states. In this work, we propose a mechanism considering the co-evolution between information states and network topology simultaneously, in which the information diffusion was executed as an $SIS$ process and network topology evolved based on the adaptive assumption. The theoretical analyses based on the Markov approach were very consistent with simulation. Both simulation results and theoretical analyses indicated that the adaptive process, in which informed individuals would rewire the links between the informed neighbors to a random non-neighbor node, can enhance information diffusion (leading to much broader spreading). In addition, we obtained that two threshold values exist for the information diffusion on adaptive networks, i.e., if the information propagation probability is less than the first threshold, information cannot diffuse and dies out immediately; if the propagation probability is between the first and second threshold, information will spread to a finite range and die out gradually; and if the propagation probability is larger than the second threshold, information will diffuse to a certain size of population in the network. These results may shed some light on understanding the co-evolution between information diffusion and network topology.
\end{abstract}


\end{frontmatter}

\section{Introduction}
\makeatletter
\newcommand{\rmnum}[1]{\romannumeral #1}
\newcommand{\Rmnum}[1]{\expandafter\@slowromancap\romannumeral #1@}
\makeatother

Recently, the popularity of the Internet has greatly facilitated information spreading, and many platforms have  emerged as tools for information spreading, such as Facebook, Twitter, and Sina Weibo \cite{Zhang2016}. Much of the previous work on information spreading has been put forward with the purpose of uncovering how information spreads on social networks and of finding ways to either enhance positive information spreading or control rumor diffusion \cite{Trpevski2010}. The main focus can be classified as having two directions.
(i) Modeling information-spreading patterns and predicting the final size of the diffusion; most of this kind of study is based on the epidemic spreading models \cite{Pastor-Satorras2015}, including the susceptible-infected-refractory (\textit{SIR}) model for rumor propagation \cite{SIR1,SIR2}, susceptible-infected-susceptible (\textit{SIS}) model \cite{SIS1,SIS2}, and susceptible-contacted-infected-refractory (\textit{SCIR}) model  \cite{SCIR1,SCIR2} for information spreading on online social media.
(ii) Designing prediction algorithms \cite{Cheng2014} according to some spreading features on the actual information spreading systems, including technology transfer \cite{Tech-Trans1,Tech-Trans2}, knowledge dissemination \cite{Know-Disse1,Know-Disse2}, and rumor propagation \cite{Rumor_Prop1,Rumor_Prop2}.\\

Despite numerous achievements in the field of information diffusion, the majority of the corresponding studies focused on information diffusion independently, in which information is transmitted from informed agents to uninformed agents through the fixed interactions between them. However, the structure of the underlying network on which information spreads always changes with time, which may influence the diffusion of information spreading significantly. To date, the adaptive behaviors, which originated in epidemic spreading  \cite{Gross2006,Sun2017}, are the most accepted assumption with which to illustrate these dynamic interactions, in which people may change their interactions in the network to protect themselves or others from being infected, with the general realization that such adaptive behaviors would suppress the diffusion process \cite{Shaw2008,Zhang2014,Liu2014a}. The case in information spreading would be more complicated; for example, one may sometimes contact or make mention to strangers with the purpose of spreading information or selling products on the social network, or one may also disconnect from the people who are spreading information on the network to prevent oneself from being disturbed \cite{Follow_Act1,Follow_Act2}; these kinds of adaptive behaviors would have an adverse impact on the spreading process. Liu and Zhang \cite{Liu2014} proposed a rewiring strategy based on the Fermi function to describe the dynamic interactions, and the simulation results indicate that this adaptive process can enhance information spreading significantly. In addition, edge-breaking \cite{Zhan2016,Zhan2018} or edge temporarily deactivating \cite{Tunc2013} were also commonly used strategies for adaptive behaviors. Although information spreading on adaptive networks has attracted much attention, the theoretical analysis of the complicated dynamic process is still not very clear.

In this paper, we concentrate on the co-evolution of information states and network topology at the same time \cite{Kozma2008,Wang2010,Guo2013}, in which information diffusion was considered an $SIS$ process and network topology evolved based on the assumption that informed individuals would rewire the informed neighbors to a random one. More importantly, we aim to provide the theoretical solution based on the Markov method to describe the coupled dynamic processes. The rest of this paper is organized as follows. In Section~\ref{Model Description}, we introduce the mechanism for network evolution and the information diffusion process. In Section~\ref{Model Analysis}, we propose the Markov-chain model to describe the mechanism mathematically, and validate the accuracy of our mathematical model with three kinds of networks. In Section~\ref{Threshold Analysis}, we aim to find the threshold values of this model. Finally, we summarize our main results and discuss some open questions for future study in Section~\ref{Conclusions}.
\section{Model Description}
\label{Model Description}
\begin{figure}[htp]
    \centerline{\includegraphics[width=10cm]{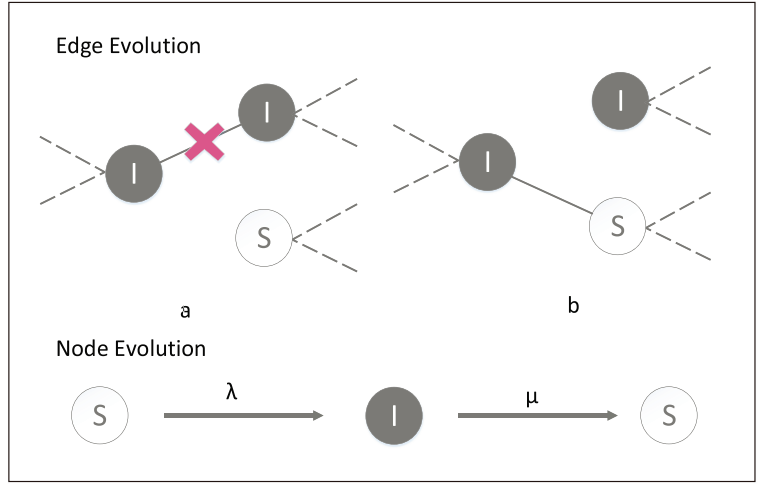}}
    \caption{\label{rewire} Diagram of the spreading model on adaptive networks. Top panel shows network evolution in which $I$-state (gray node) individuals will break the link to $I$-state neighbors (a) and reconnect to a randomly selected $S$-state individual (b). Bottom panel indicates evolution of individuals' information state on spreading process.}
\end{figure}
To explore the spreading pattern of social contagion processes on complex networks, we propose an $SIS$ model with adaptive behavior. In this model, all of the individuals in the system must be in one of two discrete states: the uninformed state (defined as $S$-state) and the informed state (defined as $I$-state) that would transmit information to their $S$-state neighbors. The model is illustrated in Figure \ref{rewire}, in which we consider two evolutional processes: the contagion dynamics and network dynamics. In the contagion dynamics, an $S$-state individual may be infected (informed) by their $I$-state neighbors with probability $\lambda$ and turn to the $I$-state. Simultaneously, an $I$-state individual may change to the $S$-state with the recovering rate $\mu$. In the network dynamics, an $I$-state individual may break the edges with their $I$-state neighbors with probability $m$ and randomly connect to a node that was not their neighbor previously. The detailed process is described as follows.

\renewcommand{\labelitemi}{$\bullet$}
\begin{itemize}
\item \textbf{Initial condition:} At the initial step, we randomly select an individual and denote it as the $I$-state and all of the other nodes as the $S$-state.
\item \textbf{Network dynamics:} At each time step, the $I$-state individuals would break the edges with their $I$-state neighbors with probability $m$, and randomly connect to a non-neighbor node.
\item \textbf{Contagion dynamics:} At each time step, the $S$-state individuals could be infected by their $I$-state neighbors with probability $\lambda$, and the $I$-state individuals could change to the $S$-state with probability $\mu$ simultaneously.
\item The steps are repeated until the number of $I$-state individuals in the network becomes stable.
\end{itemize}

\begin{table*}
\caption{\label{sym} Definitions for key parameters and variables.}
\centering
\begin{tabular}{cccc|p{4em}<{\centering}|}
\hline
$N$&number of nodes in network\\
$\lambda$&propagation rate of contagion dynamics\\
$\mu$&recovery rate of contagion dynamics\\
$m$&rewiring rate of network dynamics\\
$S_{i}(t)$($\epsilon_{i}^{S}$)&probability that node $i$ is in $S$ state at time step $t$\\
$1-S_{i}(t)$($\epsilon_{i}^{I}$)&probability that node $i$ is in $I$ state at time step $t$\\
$A_{ij}(t)$&probability that node $i$ is connected to node $j$ at time step $t$\\
$A_{ij}(0)$&adjacent matrix of network at initial step, where $A_{ij}(0)$=$A$ \\
$q_{i}(t)$&probability that node $i$ is not informed at time step $t$\\
$\Lambda_{max}(H)$&maximum eigenvalue of matrix $H$\\
\hline
\end{tabular}
\end{table*}

\section{Model Analysis}
\label{Model Analysis}
We consider a network with $N$ nodes, and the connections of the network are represented by the entries $a_{ij}$ of an $N\times N$ adjacency matrix $A$. We denote ${a_{ij}} = 1$ if node $i$ is connected with node $j$; otherwise ${a_{ij}} = 0$. In this work, we will focus on undirected, unweighted networks, indicating $a_{ij}= a_{ji}$. The main parameters used in this work are described in Table~\ref{sym}. In this section, we provide the mathematical analysis of the model using the Markov approach.

\begin{figure*}[htp]
    \centering
    \includegraphics[width=13cm]{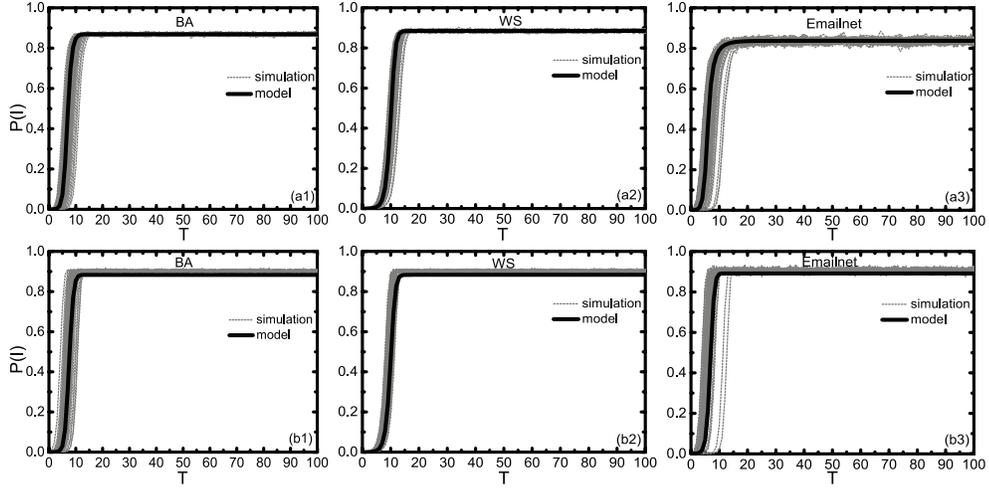}
    \caption{\label{model&sim} Evolution of fraction of $I$-state individuals both from simulation and mathematical results. (a1-a3) correspond tocase of static networks ($m=0$); (b1-b3) correspond to results for adaptive network ($m=1$). Other parameters are set as $\lambda=0.2$, $\mu=0.1$.}
\end{figure*}

\noindent{\textbf{Contagion Dynamics.}} In the contagion process, the probability of an $S$-state individual being infected only depends on the last time step according to the Markov assumption \citep{arenas2010discrete}. We can derive the probability that node $i$ is in $S$-state at time step $t$
($S_{i}(t)$) as
\begin{equation}
\label{Eq:Si}
S_{i}(t)=S_{i}(t-1)+\mu[1-S_{i}(t-1)]-S_{i}(t-1)[1-q_{i}(t-1)],
\end{equation}
where $q_{i}(t)$ is the probability that node $i$ has not been infected by any neighbor nodes at time step $t$, which can be expressed as:
\begin{equation}
\label{Eq:qi}
q_{i}(t)=\prod_{j=1}^{N}[1-\lambda A_{ij}(t)(1-S_{j}(t))].
\end{equation}
The first term of Eq.~(\ref{Eq:Si}) is the probability that node $i$ was in state $S$ at time step $t-1$, the second term represents the probability that node $i$ was in state $I$ and recovered to $S$ state at time step $t-1$, and the last term represents the probability that node $i$ was infected just at time step $t-1$. Thus, the fraction of the $I$-state nodes at time step $t$ can be given as:
\begin{equation}
\label{Eq:S}
I(t)=1-\frac{1}{N}\sum_{i=1}^N S_{i}(t).
\end{equation}

\noindent{\textbf{Network Dynamics.}} According to the network dynamics described above, the change of network structure is due to the rewiring mechanism of the $I$-state nodes in the network; thus, we only need to focus on the edges issued from the $I$-state nodes. Suppose $A_{ij}(t)$ is the probability that node $i$ is connected with node $j$ at time step $t$, where $i$ is an $I$-state node. Therefore, we can give the expression of $A_{ij}$ based on the master equation as follows:

\begin{equation}\label{Eq:Aij}
\begin{split}
A_{ij}(t)&=A_{ij}(t-1)\\
         &-A_{ij}(t-1)(1-S_{i}(t-1))(1-S_{j}(t-1))\sum_{k\neq i,j}^{N}\frac{1-A_{ik}(t-1)}{N-2}\\
         &+(1-S_{i}(t-1))\frac{1-A_{ij}(t-1)}{N-2}\sum_{k\neq i,j}^{N}(1-S_{k}(t-1))A_{ik}(t-1),
\end{split}
\end{equation}
where the first term represents the probability that link ($i$, $j$) is connected at time step $t-1$. The second term shows the decrease of the connection probability of link ($i$, $j$), which is the probability that two connected $I$-state nodes break the link at time step $t$. The third term shows the increase of the probability of link ($i$, $j$), which is the probability that two disconnected nodes (at least one $I$-state node) connect at time step $t$. In the rewiring mechanism, when a node breaks a link with their neighbor, it will connect the link with another non-neighbor node in the network. Therefore, the total number of links in the network will not change in the rewiring process. Regarding the adjacent matrix, it means that the decrease in $A_{ij}(t)$ will result in the increase of $A_{ik}(t)$ in the matrix, whereas the sum of the elements in the network remains unchanged.

In this case, the model can be described by the combination of the contagion dynamics [Eq.~(\ref{Eq:Si})] and the network dynamics [Eq.~(\ref{Eq:Aij})]. To evaluate the theoretical analysis, we applied our model to three networks.
(1) $BA$ network: $m=4$ in the BA model, where $m$ is the number of edges for the new node \cite{Barabasi1999}, and the network exhibits a power-law degree distribution $p(k)\sim k^{-\gamma}$ with $\gamma=3$.
(2) $WS$ network: rewiring each edge at random with probability $ps=1$ based on the regular network \cite{Watts1998}.
(3) $EmailNet$ \footnote{U. rovira i virgili network dataset C KONECT (Jun. 2016).}: a real-world network, which is an email network characterizing the mailing behavior between individuals.
For the sythentic networks, the network sizes are set as $N=5000$ with average degree $\langle k \rangle = 8$, and the basic statistics of the three networks are given in Table~\ref{net}. In Figure \ref{model&sim}, we show the evolution of the fraction of informed individuals both from simulation (gray dashed curves) and mathematical analysis (black solid curve). The simulation results are obtained from 100 independent realizations. From the results of the three networks, we can conclude that the mathematical approach shows good agreement with the simulation results, indicating the reasonableness of the mathematical analysis based on the Markov assumption. This conclusion is suitable for the static networks [Figure \ref{model&sim}(a1-a3), $m=0$] and adaptive networks [Figure \ref{model&sim}(b1-b3), $m=1$] simultaneously. In addition, we test the results from different values of $\lambda$ for the three networks in the \textbf{Appendix} to further illustrate the accuracy of our approach.

\begin{table}[htp]
\caption{\label{net} Basic statistics of networks, where $N$, $E$, $C$, $\langle k \rangle$, and $L$ represent the number of nodes, number of edges, clustering coefficient, mean degree, and average path length of each network, respectively.}
\centering
\begin{tabular}{cccccccc|p{4em}<{\centering}|}
\hline
  Network&$N$&$E$&$C$&$\langle k \rangle$ &$L$\\
\hline
BA&5000&19986&0.0096&7.994&3.719\\
\hline
WS&5000&20000&0.0015&8&4.377\\
\hline
Emailnet&1133&5451&0.2202&9.622&3.606\\
\hline
\end{tabular}
\end{table}

\begin{figure*}[htp]
\centering
    \includegraphics[width=12cm]{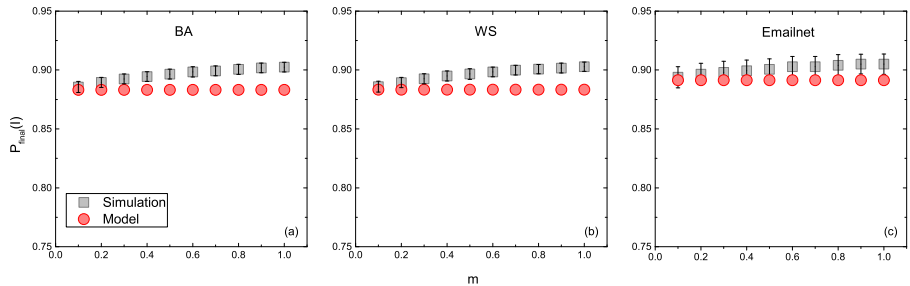}
    \caption{\label{B_I} Fraction of $I$-state individuals at the final state as a function of $m$. (a) BA network; (b) WS network; (c) Emailnet network. }
\end{figure*}

To illustrate the influence of the rewiring probability on information diffusion, we observed the fraction of the informed individuals with different rewiring probabilities. In this case, we set $\lambda=0.2$ and $\mu=0.1$, and obtain simulation and model results of the infected fraction in a steady state in Figure \ref{B_I}; the simulation results are obtained by averaging over 100 independent realizations. On each network, we find that the size of information diffusion increases with increasing rewiring probability. Therefore, the rewiring mechanism can promote information spreading on different networks, and it would be a reasonable strategy to enhance the information spreading. It should be noted that a larger deviation between simulation and mathematical analysis is observed when $m$ becomes larger. The possible reason would be that we use the connection probability to illustrate the network structure in the mathematical analysis, while the edges between two nodes only exist or do not in the simulation. Thus, the difference would emerge when the rewiring probability is large enough.
\section{Threshold Analysis}
\label{Threshold Analysis}
\begin{figure}[htp]
  \centerline{\includegraphics[width=10cm]{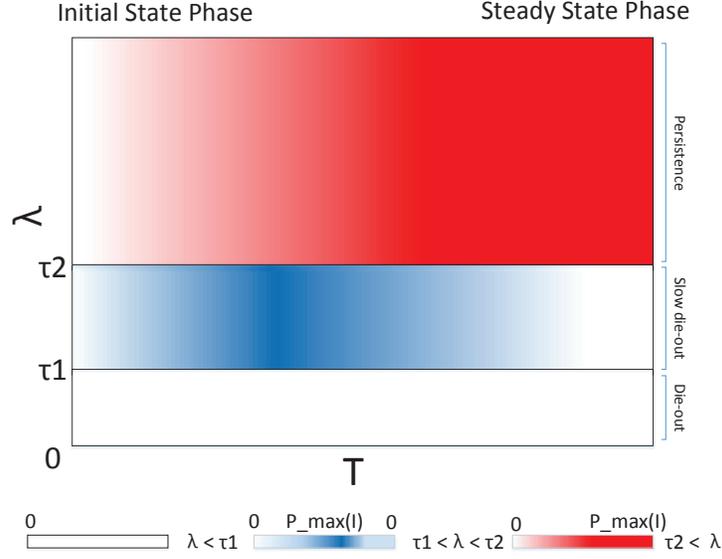}}
  \caption{\label{tau}Schematic of information evolution under influence of $\lambda$.}
\end{figure}

When the information is spreading among a population, one of the most important things to know is whether the information will break out or not, which indicates the threshold value of the corresponding dynamics. Since the network we are concerned with is an adaptive network, the threshold value of the information spreading on this network becomes more complicated than that on the static networks. Therefore, we first analyze the threshold value of information spreading on a static network (i.e., the case $m=0$). Assuming $\lambda_{c}$ is the critical value of the information transmission rate for fixed values of $\mu$, when $\lambda$ $<$ $\lambda_{c}$, the final fraction of informed individuals is $I_{final}$=0. If $\lambda$ $>$ $\lambda_{c}$, there would be a part of the population that would be informed in the system, indicating $I_{final}$$>0$. Letting the probability that node $i$ is in the $I$-state at time step $t$, $\epsilon_{i}^{I}=1-S_{i}(t-1)$, when $\lambda$ $\longrightarrow$ $\lambda_{c}$, we have $\epsilon_{i}^{I}\approx0$, and thus we can obtain the probability that node $i$ has not been informed at time step $t$ [denoted $q_{i}(t)$] as follows:
\begin{equation}
\label{Eq:qi1}
\begin{split}
q_{i}(t)&=\prod_{j=1}^{N}[1-\lambda A_{ij}(t)(1-S_{j}(t))]\\
        &=\prod_{j=1}^{N}[1-\lambda A_{ij}(t)\epsilon_{j}^{I}]\\
        &\approx 1-\lambda\sum_{j} A_{ij}(t)\epsilon_{j}^{I}.
\end{split}
\end{equation}
Substituting Eq.~(\ref{Eq:qi1}) into Eq.~(\ref{Eq:Si}), we obtain
\begin{equation}
\mu\epsilon_{i}^{I}=\epsilon_{i}^{S}[\lambda\sum_{j} A_{ij}(t-1)\epsilon_{j}^{I}].
\end{equation}
Furthermore, we can obtain
\begin{equation}
(\epsilon_{i}^{S}A_{ij}-\frac{\mu}{\lambda}E)\epsilon_{j}^{I}=0,
\end{equation}
where $E$ is the identity matrix. Accordingly, the threshold value is $\displaystyle\lambda_{c}=\frac{\mu}{\Lambda_{max}(H)}$, and the element in matrix $H$ is $\epsilon_{i}^{S}A_{ij}$. At the critical point, we have $\epsilon_{i}^{S}\approx 1$, and thus $H\approx A_{ij}$, where $\Lambda_{max}(H)$ is the maximum eigenvalue of $H$ \cite{Guerra2010, Trajanovski2015,Sayama2013,Chandrasekar2014,Belykh2014}.

Therefore, the threshold value is $\displaystyle\lambda_{c1}=\frac{\mu}{\Lambda_{max}(A)}$ when $m=0$ (static network), which is defined as \textbf{the first threshold value of the model} ($\tau1$ in Figure~\ref{tau}). However, the network is changing with the rewiring mechanism designed in our model, leading to the constantly changing threshold value. To guarantee that the information can spread out on the adaptive network, we must find the maximum value of $\displaystyle\lambda_{c1}=\frac{\mu}{\Lambda_{max}(A)}$ [i.e., the minimum value of $\Lambda_{max}(A)$]. As long as $\lambda > max\{\lambda_{c1}\}$, the information can break out. We define $\lambda_{c2} \doteq max\{\lambda_{c1}\}$ as \textbf{the second threshold value of the model} ($\tau2$ in Figure~\ref{tau}). To calculate $\lambda_{c2}$, we first introduce a theorem as follows:

\begin{theorem} [Gershgorin circle theorem \cite{bauer1968fields}]
Let $A$ be a complex $N\times N$ matrix, with entries $a_{ij}$. Every eigenvalue $\Delta$ of $A$ must lie within at least one of the following discs:
\end{theorem}
\begin{equation}
\mid\Delta-a_{ii} \mid \leq r_{i}=\sum_{j=1,j\neq i}^{N}|a_{ij}|,  i=1,2,...,N.
\end{equation}

With the evolution of the network structure, the value of the elements in matrix $A(t)$ (except the diagonal elements value) will lie between 0 and 1 (not equal to 0 or 1). To guarantee that information can spread out, the originating propagation rate should be greater than the maximum value of the threshold value. In addition, the upper limit of the second threshold value can be estimated by the minimum value of the maximum eigenvalue for matrix $A(t)$. According to the \textit{Gershgorin circle theorem}, the minimum value of the maximum eigenvalue would be obtained when the adjacent matrix satisfies the following conditions: (a) the main diagonal elements of matrix $A$ are zero, and (b) all of the row sums of matrix $A$ are equal to the same value. It is obvious that $\displaystyle\max(r_{i})\geq\frac{r_{1}+r_{2}+...+r_{n}}{n}$, and only if $r_{1}=r_{2}=...=r_{n}$ does $\max(r_{i})$ take the minimum value $\displaystyle\frac{r_{1}+r_{2}+...+r_{n}}{n}$ (the equality holds) according to mean inequality \cite{Clarke1994}. We use $Z$ to express the adjacent matrix $A$ when it satisfies conditions (a) and (b). Therefore, the second threshold value of the model is $\displaystyle\lambda_{c2}=\frac{\mu}{\Lambda_{max}(Z)}$. When the transmission probability $\lambda > \lambda_{c2}$, the information can always spread out to a certain number of individuals.

Figure \ref{tau} is a schematic of the relationship between the thresholds and the fraction of $I$-state individuals. According to the analysis above, we can conclude that when $\lambda < \lambda_{c1}$ the information cannot spread out (the bottom part of Figure \ref{tau}, i.e., the white area). When $\lambda_{c1} < \lambda < \lambda_{c2}$ (the middle part of Figure \ref{tau}, i.e., the blue area), the information will first spread to a number of individuals, and then the fraction of $I$-state individuals will tend to be zero in the final state, which is a case called "slow information die-out." The top part of Figure \ref{tau} (red area) is a case of information persistence, which means that there will be a number of individuals known about the information in the steady state when $\lambda > \lambda_{c2}$. In this case, we can give the threshold values of the model from a theoretical point, i.e., the first threshold value $\lambda_{c1}$ and the second threshold value $\lambda_{c2}$. According to the theoretical computation method given above, we calculate the threshold values for the information spreading on the three networks in Table~\ref{threshold}.

\begin{figure*}[htp]
\centering
    \includegraphics[width=13.5cm]{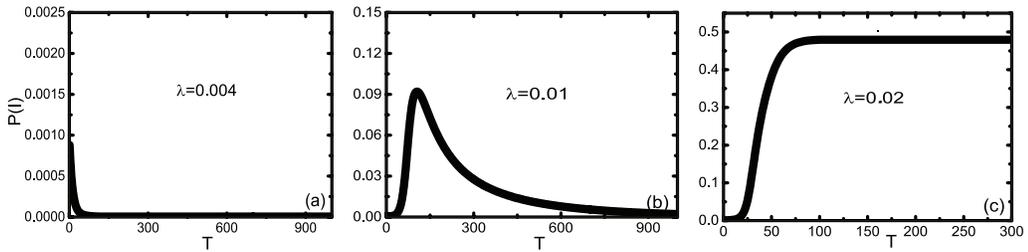}
    \caption{\label{fig5}Evolution of fraction of $I$-state individuals under different values of $\lambda$: (a) $\lambda=0.004$; (b)$\lambda=0.01$; (b)$\lambda=0.02$.}
\end{figure*}

\begin{figure*}[htp]
\centering
    \includegraphics[width=12.5cm]{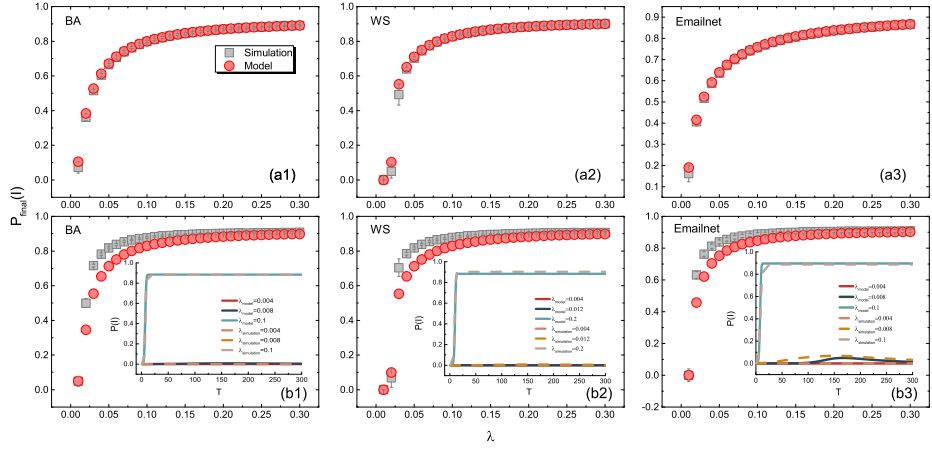}
    \caption{\label{B&I} Fraction of $I$-state individuals at final state as a function of $\lambda$. (a1-a3) Static networks($m=0$); (b1-b3) adaptive networks ($m=1$), with the inset showing three different situations: $\lambda<\lambda_{c1}$, $\lambda_{c1}<\lambda<\lambda_{c2}$, and $\lambda<\lambda_{c2}$.}
\end{figure*}

Figure \ref{fig5} is a schematic of information evolution under different values of transmission rate $\lambda$, which displays three different dynamical behaviors, i.e., information die-out, slow information die-out, and information persistence. We verify these results on the $Emailnet$ network in Figure \ref{fig5} by choosing different values of $\lambda$. When $\lambda=0.004$, which is smaller than the first threshold value of this network (i.e., 0.00482), there is no information spreading in the population [Figure \ref{fig5}(a)]. However, the fraction of $I$-state individuals will first reach a small peak and then decrease to zero when $\lambda=0.01$ ($\lambda_{c1}<\lambda<\lambda_{c2}$), corresponding to Figure \ref{fig5}(b) (slow information die-out). When $\lambda=0.02$ ($\lambda>\lambda_{c2}$), the evolution of the fraction of infected individuals is an S-shaped curve, corresponding to Figure \ref{fig5}(c) (information persistence). Figure \ref{B&I} shows the final size of the population of informed individuals as a function of $\lambda$ both for the static networks [$m=0$, Figure \ref{B&I}(a1-a3)] and adaptive networks [$m=1$, Figs. \ref{B&I}(b1-b3), the inset of which shows three different situations: (1)$\lambda<\lambda_{c1}$, (2)$\lambda_{c1}<\lambda<\lambda_{c2}$, and (3)$\lambda<\lambda_{c2}$). The simulation results are consistent with the mathematical approach in both cases, indicating the reasonableness of the mathematical analysis.

\begin{table}[htp]
\caption{\label{threshold}Diffusion threshold values for the three studied networks.}
\centering
\begin{tabular}{cccc|p{4em}<{\centering}|}
\hline
  Network&$\lambda_{c1}$&$\lambda_{c2}$\\
\hline
BA&0.00480&0.01251\\
\hline
WS&0.01171&0.01250\\
\hline
Emailnet&0.00482&0.01039\\
\hline
\end{tabular}
\end{table}

\section{Conclusions \& Discussion}
\label{Conclusions}
Aiming to give a better understanding of information diffusion on adaptive networks, in this work, we built an $SIS$ model considering the co-evolution of information states and network topology at the same time. We presented mathematical analyses to illustrate the co-evolution dynamics according to the Markov approach, and the results of both simulation and mathematical analyses show good agreement on three different networks (the BA network, WS network, and a real-world email network). According to the mathematical analyses, we found that there are two threshold values ($\lambda_{c1}$ and $\lambda_{c2}$) for this model, which is different from previous studies. We validated our results by simulations using the three different networks and found that, when the spreading probability $\lambda<\lambda_{c1}$, information cannot diffuse in  thesystem; when $\lambda_{c1}<\lambda<\lambda_{c2}$, information will first propagate to a certain number of the population and gradually become extinct; and when $\lambda>\lambda_{c2}$, there will always be a certain number of the population that knows about the information.

Furthermore, we observed that information spreading with an adaptive process can increase the informed popularity. This induces us to pay more attention to information spreading on the dynamical social network, which may infect a large number of the population with adverse impact, e.g., the salt-buying panic in China caused by an earthquake in Japan and the Fukushima reactor meltdown in 2011 \cite{LiuX2011}. In a planned future study, more detailed data about information spreading on adaptive networks will be needed to achieve in-depth understanding of the dynamics of information spreading. In conclusion, this research enhances understanding of information spreading on adaptive networks.

%
%
%
%
%
%
%

\section{Acknowledgments}
This work was partially supported by the Zhejiang Provincial Natural Science Foundation of China (Grant Nos. LR18A050001 and LY18A050004), the National Natural Science Foundation of China (Grant Nos. 61673151, 61873080 and 11671241), the Major Project of The National Social Science Fund of China (Grant No. 19ZDA324),  and Outstanding Young Talents Support Plan of Shanxi province, Selective Financial Support for Scientific and Technological Activities of Overseas Students in Shanxi Province.

\end{document}